\documentclass{article}

\usepackage{arxiv}

\usepackage[utf8]{inputenc} 
\usepackage[T1]{fontenc}    
\usepackage{hyperref}       
\usepackage{url}            
\usepackage{booktabs}       
\usepackage{amsfonts}       
\usepackage{nicefrac}       
\usepackage{microtype}      
\usepackage{lipsum}
\usepackage{graphicx}
\usepackage{float}
\usepackage{mathtools}
\usepackage[english]{babel}

\newtheorem{theorem}{Theorem}

\title{No Substitute for Functionalism - A Reply to `Falsification \& Consciousness'}

\author{
  Natesh Ganesh \\
  Information Technology Lab, ACMD, NIST Boulder\\
  Dept of Physics, University of Colorado, Boulder\\
  Boulder, Colorado 80305 \\
  \texttt{natesh.ganesh@colorado.edu} \\
}

\begin{document}
\maketitle

\begin{abstract}
 In their paper 'Falsification and Consciousness' \cite{Hoel}, Kleiner and Hoel introduced a formal mathematical model of the process of generating observable data from experiments and using that data to generate inferences and predictions onto an experience space. The resulting \textit{substitution argument} built on this framework was used to show that any theory of consciousness with \textit{independent} inference and prediction data are pre-falsified, if the inference reports are considered valid. If this argument does indeed pre-falsify many of the leading theories of consciousness, it would indicate a fundamental problem affecting the field of consciousness as a whole that would require radical changes to how consciousness science is performed. In this reply, the author will identify avenues of expansion for the model proposed in \cite{Hoel} allowing us to distinguish between different types of variation. Motivated by examples from neural networks, state machines and Turing machines, we will prove that substitutions do not exist for a very broad class of \textit{Level-1} functionalist theories, rendering them immune to the aforementioned substitution argument.
\end{abstract}


\section{Introduction}
A formal model of generating data through experiments in consciousness science and then using this data to make \textit{predictions} and \textit{inferences} onto an experience space was introduced in \cite{Hoel}. This model was then used to propose a clear definition of falsification, followed by the\emph{`substitution arguments'}. The authors also pointed out that the unfolding argument from \cite{Doerig} would be a special case of their results. It is very interesting work, accessible and proposes a necessary descriptive mathematical framework that could prove to be very useful moving forward. The author will assume that the readers are familiar with the work in \cite{Hoel} and for the sake of clarity, we will try and borrow the symbols and terminologies from it as much as possible.

The reply is structured as follows - we will start by discussing the main definitions and theorem(s) from \cite{Hoel} in section 2, identify and correct specific aspects of the model from the original work. Following this expansion, we will then use some of the proposed examples for substitutions in \cite{Hoel} to show why these specific substitutions do not imply pre-falsifications for a broad class of Level-1 functionalist theories in section 3. In section 4, we will follow it up with a formal definition of this class of functionalist framework and a proof that no substitutions exist for them. In section 5, we will briefly explore the difference between Level-1 and Level-2 functionalism. The note will conclude in section 6 summarizing the work and briefly discussing the implications.

\section{Understanding the Substitution Argument}
\begin{figure} 
\begin{center}
\includegraphics[scale=0.65]{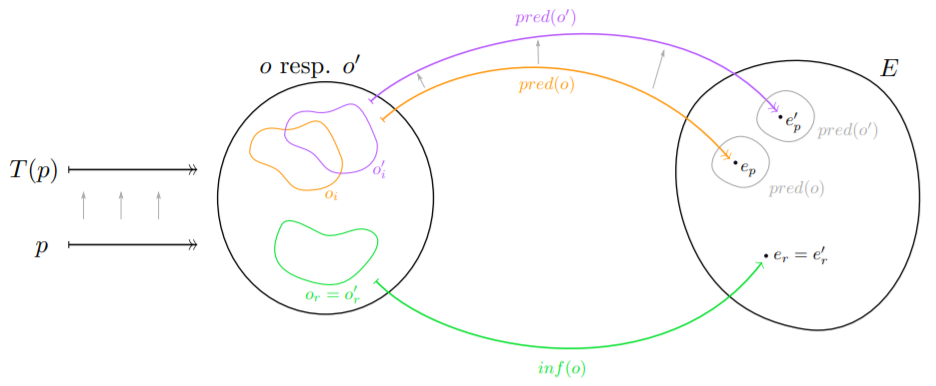}
\end{center}
\caption{Borrowed from \cite{Hoel} - `This picture illustrates substitutions. Assume that some data set $o$ with inference content $o_r$ is given. A substitution is a transformation $T$ of physical systems which leaves the inference content or invariant but which changes the result of the prediction process. Thus whereas $p$ and $T(p)$ have the same inference content or, the prediction content of experimental data sets is different. Different in fact to such an extent that the predictions of consciousness based on these datasets are incompatible (illustrated by the non-overlapping circles on the right). Here we have used that by definition of $P_{o_r}$, every $p \in  P_{o_r}$ yields at least one data set $o^{\prime}$ with the same inference content as $o$ and have identified as $o$ and $o^{\prime}$ in the drawing.'}
\label{Hoel_figure}
\end{figure}

We start with one of the central definitions and results from \cite{Hoel} (refer to Fig.(\ref{Hoel_figure}) from \cite{Hoel}) that will be the focal point of the discussions here. These include -
\begin{itemize}
\item[(a)] \textbf{Definition 2.1} - \emph{falsification} is defined as \emph{`there is a falsification at $o \in O$ if we have $inf(o) \notin pred(o)$'}. 
\item[(b)] \textbf{Definition 3.1} - $o_r$\emph{-substitution} is defined  as \emph{`a $o_r$-substitution if there is a transformation $S : P_{o_r} \rightarrow P_{o_r}$ such that at least for one $p \in P_{o_r}$ -  $pred \cdot obs(p) \cap pred \cdot obs(S(p)) = \phi$'}.
\item[(c)] \textbf{Definition 3.8} - Inference and prediction data is defined as \textit{independent} \textit{`if for any $o_i$, $o_i^{\prime}$ and $o_r$, there is a variation $v: P \rightarrow P$ such that $o_i \in obs(p)$, $o_i^{\prime} \in  obs(v(p))$, but $o_r \in obs(p)$ and $o_r \in obs(v(p))$ for some $p \in P$.'}
\item[(d)] In section (3.4.1), \textit{minimally informative} is defined as \textit{that for every} $o \in O$,\textit{ there exists an} $o^{\prime} \in O$ such that $pred(\bar{o}) \cap pred(\bar{o}^{\prime})= \emptyset$.
\item[(e)] The substitution argument is given in \textbf{Theorem 3.10} - \textit{`If inference and prediction data are independent, either every single inference operation is wrong or the theory under consideration is already falsified.'} 
\end{itemize}

One of the major things to note is that while the \textit{minimally informative} criterion guarantees that there are at least two sets of observable data with different predictions, it does not constrain their corresponding inferences in any manner \cite{Jake}. Thus frameworks for which we have $pred(\bar{o}) \cap pred(\bar{o}^{\prime})= \emptyset$ could also have $inf(o) \cap inf(o^{\prime})= \emptyset$ and still be \textit{minimally informative}. A more robust and expanded definition of independence can be built based on this observation by expanding on what definition (3.8) intended to capture - \textit{``in most experiments, the prediction content $o_i$ and inference content $o_r$ consist of different parts of a dataset. What is more, they are usually assumed to be independent, in the sense that changes in $o_i$ are possible while keeping $o_r$ constant''} \cite{Hoel}. While we can have ($o_i, o_i^{\prime}$) pairs that preserve $o_r$, it does not necessarily imply that all ($o_i, o_i^{\prime}$) pairs have to satisfy this constraint (and it would be erroneous to assume otherwise). Thus we could have a set of prediction data $\{ o_i^{\prime} \}$ generated by a \textit{variation} (as defined in \cite{Hoel}) that maintains the same $o_r$, but does not contain $o_i^{\prime \prime}$ (or a set of them) with $inf(o) \cap inf(o^{\prime \prime})= \emptyset$ that ensures that the minimally informative criterion ($pred(\bar{o}) \cap pred(\bar{o}^{\prime \prime})= \emptyset$ ) is met. Furthermore varying $o_i$ while keeping $o_r$ constant does not necessarily mean we vary the prediction $pred(o_i)$ i.e. we can have $o_i \neq o_i^{\prime}$ but still have $pred(o_i) = pred(o_i^{\prime})$. This expanded idea of independence is aligned with the underlying motivation, while not being equivalent to falsification by definition \cite{Jake}.

Utilizing the original definition of \textit{variation}, we see that for any ($o_i, o_i^{\prime}$) pair that maintain the same $o_r$ there are  variations of two types -
\begin{itemize}
    \item [(i)] \textit{Type-1 variations} where $o_i \neq o_i^{\prime}$, but $pred(o) \cap pred(o^{\prime}) \neq \emptyset$.
    \item [(ii)] \textit{Type-2 variations} where $o_i \neq o_i^{\prime}$, but $pred(o) \cap pred(o^{\prime}) = \emptyset$.
\end{itemize}

It is immediately evident that while both variations allow for prediction and inference data to be independent, only Type-2 variations imply pre-falsifications. Thus independence (under this expanded set of variations) does not necessarily imply falsifications by definition and we could have frameworks with prediction and inference data independence through only Type-1 variations. Thus \textbf{Theorem 3.10} is incomplete at best since \textbf{Definition 3.8} of \emph{independence} does not account for \textit{Type-1 variations}. The correct restatement of the theorem would then be -

\begin{theorem}
\textit{`If inference and prediction data are independent under a \textbf{Type-2 variation}, either every single inference operation is wrong or the theory under consideration is already falsified.’}
\end{theorem}

Since all variations that produce independence do not immediately entail falsification, the claim that - \textit{'our results show that if the independence of prediction and inference holds true, as in contemporary cases where report about experiences is relied upon, it is likely that no current theory of consciousness is correct'} in \cite{Hoel} only holds if is shown that the variation is of Type-2 for the theory. The question now becomes one of determining whether substitutions identified in different existing frameworks are produced by variations of Type-1 or 2. This will be the focus of the next section with respect to some of cases suggested in the paper. 

The authors of \cite{Hoel} also make the case for why $obs: P \rightarrow O$ and $pred: O \rightarrow E$ mappings are better characterized as \emph{correspondences}, while $inf$ can be considered to be a function.  They argue that while $inf$ can be a correspondence, it can redefined to be a function by redefining the experience space $E$ to a new space $E^{\prime} \coloneqq \{inf(o)|o \in O\}$, \textit{'where every individual element} $e^{\prime} \in E^{\prime}$ \textit{describes what can be inferred from one dataset} $o \in O$, \textit{so that} $inf^{\prime} \coloneqq O \rightarrow E^{\prime}$ \textit{is a function.'} While this might be coherent mathematically, the authors do not explain the effect of redefining $E$ to $E^{\prime}$ has on the $pred$ correspondence or explore the more general case of both $pred$ and $inf$ being correspondences (if it does not affect the actual theorems). If $inf(o)$ is a subset of elements in $E$ as opposed to a single element, it would entail a change in falsification to $inf(o) \cap pred(o) = \emptyset$ and complicate both variations and substitutions since we could now have variations that maintain $o_r$ and have $pred(\bar{o}) \cap pred(\bar{o}^{\prime})$ but still not entail falsification since $pred(o) \cap inf(o) \neq \emptyset$ and $pred(o^{\prime}) \cap inf(o^{\prime}) \neq \emptyset$. However this is a minor objection and unrelated to the rest of this reply. Since this is a first step in formalizing these ideas, it is justifiable for the authors of \cite{Hoel} to not cover the more general cases in their work yet. However one hopes that the change in $inf$ from correspondence to function is explored in detail and if it does not affect the final results as claimed in the paper, a clear and rigorous derivation of that will be provided in future work.

\section{Application of the Substitution Argument to Neural Networks \& Turing Machines}
The authors of \cite{Hoel} would have served the readers better by providing a very clear example of the application of the substitution argument to some of the proposed cases of data independence stated in the paper. In this section, we will examine the existence and type of variations in the case of powerful and widely used models of computation - neural networks, a simple finite state machines and Turing machines. Let us start with the case of artificial neural networks discussed in the paper. The authors state - \textit{`For any ANN, report (output) is a function of node states. Crucially, this function is non-injective, i.e., some nodes are not part of the output. For example, in deep learning, the report is typically taken to consist of the last layer of the ANN, while the hidden layers are not taken to be part of the output. Correspondingly, for any given inference data, one can construct a ANN with arbitrary prediction data by adding nodes, changing connections and changing those nodes which are not part of the output. Put differently, one can always substitute a given ANN with another with different internal observables but identical or near-identical reports. From a mathematical perspective, it is well-known that both feedforward ANNs and recurrent ANNs can approximate any given function (Hornik et al. 1989; Schafer and Zimmermann 2007). Since reports are just some function, it follows that there are viable universal substitutions.'} It seems like the authors are claiming that the arguments presented in their discussion imply the existence of \emph{universal substitutions} for neural networks but is that really the case? Since there is no example accompanying the claim, we will take up a simple example of a recurrent neural network and apply the substitution argument as faithfully as possible to the discussion above. We would like to acknowledge that a bulk of the work in this section is not very novel and is built on the analysis performed by Hanson and Walker in \cite{Walker2} and \cite{Walker}. 

Consider a general recurrent neural network (RNN) $\mathcal{N}$ with $\mathcal{I}$ number of input neurons, an arbitrary number $\mathcal{H}$ of hidden neurons and $\mathcal{T}$ output neurons. Let the input-to-hidden neuron weights, hidden-to-hidden weights and hidden-to-output weights be given by $W_\mathcal{I}$, $W_\mathcal{H}$ and $W_\mathcal{T}$. The dynamics of the hidden states $h_t$ and output $o_t$ of the recurrent neural network at time $t$ is given as
\begin{eqnarray}\label{RNN}
    h_t &=& \sigma (W_\mathcal{H} h_{t-1}+W_\mathcal{I} x_t) \nonumber \\
    o_t &=& W_\mathcal{T} h_{t}
\end{eqnarray}

where $\sigma$ can be a non-linear activation function like ReLu or sigmoid. Note that $h_t$ is a $\mathcal{H}$-dimensional vector. By the discussion of such neural networks in \cite{Hoel}, we have the hidden states $h_t$ as part of the prediction data and output states $o_t$ as the inference data. From existing literature we know that it is possible to construct a different RNN $\mathcal{N}^{\prime}$ that can produce the same input-output behavior but with a different hidden-to-hidden weight matrix. Thus $\mathcal{N}^{\prime}$ would have the same number of input and output neurons $\mathcal{I}$ and $\mathcal{T}$ respectively as $\mathcal{N}$. However we assume that $\mathcal{N}^{\prime}$ has a different number of hidden neurons $\mathcal{H}^{\prime}$. The corresponding weight matrices are given as $W_\mathcal{I}^{\prime}$, $W_\mathcal{H}^{\prime}$ and $W_\mathcal{T}^{\prime}$. We will assume that the new hidden state of this RNN is given as $h_t^{\prime}$ (a $\mathcal{H}^{\prime}$-dimensional vector) while inputs and outputs remain the same at $x_t$ and $o_t$. The dynamics are given as

\begin{eqnarray}\label{RNN_eqn}
    h_t^{\prime} &=& \sigma (W_\mathcal{H}^{\prime} h_{t-1}^{\prime}+W_\mathcal{I}^{\prime} x_t) \nonumber \\
    o_t &=& W_\mathcal{T}^{\prime} h_{t}^{\prime}
\end{eqnarray}


Thus $\mathcal{N}$ and $\mathcal{N}^{\prime}$ both give the same output $o_t$ (corresponding to inference data) while changing the hidden states (corresponding to prediction data).  $h_t \neq h_t^{\prime}$ since they are vectors of different dimensions. Under the definitions used in \cite{Hoel}, this would imply that prediction and inference data are independent and viable substitutions exist. But whether this particular substitution in the neural network implies falsification depends on whether or not the variation is of Type-1 or 2. The answer is that it depends upon the framework or theory of interest and it's effect on the $pred$ function.

We will explore this further with a functionalist framework $\mathcal{F}$. We define a functionalist framework to be one in which - \textit{`the states are typically described in terms of functional behaviors ("stop", "walk", "go", etc.) but what really gives them meaning mathematically is only their topological relationship with one another. This implies that at this level, the formal description of the computation is not grounded in any particular physical representation and could, in fact, be realized by radically different causal structures. This abstract treatment of computation corresponds to what Chalmers’ refers to as the ``finite-state automaton'' (FSA) level of description, due to the fact it is defined in terms of a global finite-state automaton. Beneath this level is what Chalmers refers to as the "combinatorial-state automaton" (CSA) description. The only difference between the FSA and CSA levels of description is that the latter specifies the computational states of the former in terms of a specific labeling or encoding of the subsystems that comprise the global system.'} \cite{Walker}, \cite{Chalmers}. We will simply refer to the FSA and CSA levels of description as \textbf{Level-1} and \textbf{Level-2} descriptions respectively in order to be more general and avoid any baggage with the terms - FSA and CSA. 

Under this computational hierarchy, we have $\mathcal{F}$ to be a Level-1 functionalist framework. While specific representations are important to understand the physical implementation, the $pred$ function for such a $\mathcal{F}$ would be dependent on the functional states $\{s \}$ as defined in the Level-1 description only and not on the particular of any specific encoding at Level-2 i.e. Level-1 functionalist frameworks are not representationalist simply because they employ representation. In the case of the neural network, the states in the Level-2 description are defined by their relationship to other states $\{ h_{t-1}, x_t\} \rightarrow \{h_t, o_t\}$ i.e. their functional structure. Though $h$ and $h^{\prime}$ are vectors of different dimensions (corresponding to their particular physical representations in their respective networks), both networks maintain the same input-output relationship and as a result, we would have for every $h \in \{h\}$ a corresponding $h^{\prime} \in \{ h^{\prime} \}$ such that $\{ h_{t-1}, x_t\} \rightarrow \{h_t, o_t\}$ and $\{ h_{t-1}^{\prime}, x_t\} \rightarrow \{h_t^{\prime}, o_t\}$ for all input-output pairs $(x_t,o_t)$. Thus $h_t$ and $h_t^{\prime}$ correspond to the same functional state $s_t$ (say) at Level-1. We can think of the $pred$ function for $\mathcal{F}$ as composed of two functions - $abs$ which maps the particular physical realization of Level-2 to the higher Level-1 FSA description, followed by a $pred^{\prime}$ function that maps the (abstract) functional state $s_t$ to the experience space - $pred()=pred^{\prime}(abs())$. Hence we have the specific encodings of the hidden states $h_t \neq h_t^{\prime}$ but would still have $pred(h_t)=pred(h_t^{\prime})=pred^{\prime}(s_t)$ (since $pred$ function only depends upon the Level-1 functional state), while the output remains the same for both $\mathcal{N}$ and $\mathcal{N}^{\prime}$. Thus the variation from $\mathcal{N}$ and $\mathcal{N}^{\prime}$ does not imply pre-falsification. It is trivial to see that if we were drop the time element $t$ from the states, the same arguments can be extended to feedforward neural networks with varying number of hidden neurons. Since the assumptions made here are very general, we claim that \textbf{for a Level-1 functional framework $\mathcal{F}$, any substitution of \textit{`a given ANN with another with different internal observables but identical or near-identical reports'} \cite{Hoel} will always correspond to a Type-1 variation and never imply pre-falsification of $\mathcal{F}$ if we take the inferences to be true.}

To see the dependence of the above discussion on the framework in question,  let us consider a different framework of consciousness $\mathcal{F}^{\prime}$ in which the result of the prediction function is dependent on the dimensionality of the hidden-state vector i.e. $pred(h_t)=g(dim(h_t))$. In this framework we can clearly see that $pred(h_t)$ need not be equal to $pred(h_t^{\prime})$ since the two vectors are of different dimensions. In such case, the relationship between $N$ and $N^{\prime}$ corresponds to a type-2 variation and we would have framework $\mathcal{F}^{\prime}$ to be falsified by Theorem 1. While the dependence on dimensionality of the hidden state is a contrived example that is easy to visualize, the same argument would apply to any framework that distinguished between any $h_t$ and $h_t^{\prime}$ (that have the same functional structure as per a Level-1 description) with respect to the $pred$ function as a result of the difference in their physical encoding/representation i,e, CSA or Level-2 description. We will see this in greater detail when studying an example comparing systems that do and do not contain any feedback next.

Diving deeper into the differences between Level-1 (FSA) and Level-2 (CSA) descriptions - \textit{``In digital electronics, as well as models of the human brain, this encoding is usually given in terms of binary labels that are assigned to instantiate the functional states of the system. Consequently, transitions between states in the CSA description fix local dependencies between elements, as the correct Boolean update must be applied to each ``bit'' or ``neuron'' based on the global state of the system. Furthermore, once a binary representation is specified it constrains the memory required to instantiate the computation, as the number of bits that comprise the system is now fixed. The final level of the hierarchy is the specific choice of logic gates used to implement the Boolean functions specified at the CSA level.} \cite{Walker}. This is best understood from the example worked out in \cite{Walker} where they show that a causal structure theory of consciousness like Integrated Information Theory (IIT) makes different predictions for different CSA representations and as a result are either pre-falsified or are unfalsifiable depending on whether the inference is made at the FSA/Level-1 or CSA/Level-2 respectively.

\begin{figure} 
\begin{center}
\includegraphics[scale=0.5]{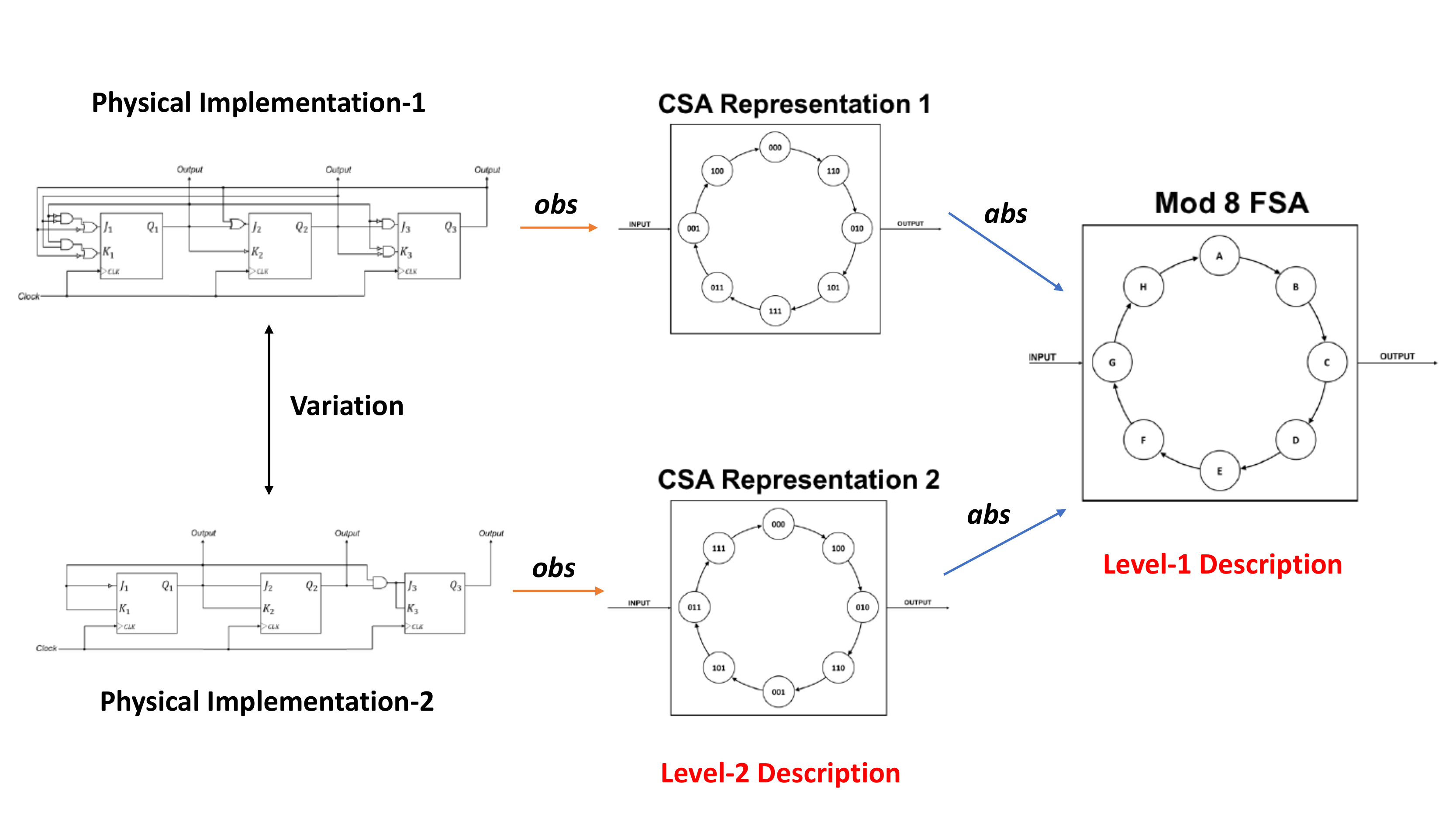}
\end{center}
\caption{Modified from \cite{Walker} - We created the mirror image of the Fig.(1) from \cite{Walker} by starting from two different physical implementations 1 and 2 of the tollbooth on the left, having feedback \& no feedback respectively. This picture is closer to the idea of performing experiments and collecting data from a physical system. Since both implementations maintain the same input-output relationship while producing different representations of their internal states, they are \textit{variations} of each other. The corresponding  CSA encoding representations that constitute the Level-2 description are given in the middle. We can see that an abstraction function $abs$ will generate the same higher Level-1 description of the Mod 8 FSA in both cases.}
\label{Walker_fig}
\end{figure}

We will borrow Fig.(1) from \cite{Walker} and modify it to fit the case here to give Fig.(\ref{Walker_fig}). We see that starting with the two physical implementations 1 and 2, we have two physical systems from which our prediction and inference data is generated using the $obs$ correspondence. We will have the internal states and outputs of the implementation correspond to prediction and inference data respectively. Since both implementations maintain the same output while having different CSA encoding representations for the internal states, the two systems can be seen as a variation of each other. This difference in the state representations correspond to the difference in their physical implementations - feedback vs no-feedback. We can then generate the higher Level-1 FSA description that represents the functional structure shared by both Level-2 CSA descriptions using a suitable abstraction functions $abs$. From the Fig.(\ref{Walker_fig}), we have the state $100^1$ from CSA representation-1 to have the same topological relationships as the state $111^2$ from CSA representation-2, and both of these correspond to state $H$ in the Level-1 description. Since the predictions of a Level-1 functionalist framework only depend on functional states in the Level-1 description, we would have $pred(100^1)=pred(111^2)=pred^{\prime}(H)$. Thus variations like the one in Fig.(\ref{Walker_fig}) that maintain the input-output relationship also maintain the functional structure and are thus Type-1 variations that do not pre-falsify the framework. On the other hand, we can see that for frameworks that produce different predictions based on different CSA representations i.e. $pred(100^1) \neq pred(111^2)$, such a variation would be of Type-2 and imply pre-falsification if the inferences at the Level-1 description are considered to be valid. 

A similar claim on the existence of substitutions for Turing machine was also made - \textit{`Turing machines are extremely different in architecture than ANNs. Since they are capable of universal computation, they should provide an ideal candidate for a universal substitution'} \cite{Hoel}. To explore this, we will start with a mathematical definition of a Turing machine $\mathcal{M}$ from \cite{Hopcroft} as a 7-tuple $\mathcal{M}=\{ Q, \Lambda, b, \Sigma, \delta, q_0, F\}$ where

\begin{itemize}
    \item $Q$ is a finite, non-empty set of states.
    \item $\Lambda$ is a finite, non-empty set of tape alphabet symbols.
    \item $b \in \Lambda$ is the blank symbol.
    \item $\Sigma \subseteq \Lambda \backslash \{b\} $ is the set of input symbols.
    \item $q_0 \in Q$ is the initial state.
    \item $F \subseteq Q$ is the set of final or accepting states.
    \item $\delta : (Q \backslash F) \times \Lambda \rightarrow Q \times \Lambda \times \{L,R\}$ is a partial function called the transition function.
\end{itemize}

In a Turing machine, the input is encoded into the tape and Turing machine (or specifically the state registry) starts at an initial starting state. There is a transition function that maps the current state of the state registry and the symbol on the tape to a new state of the registry, a change in the symbol on the tape at the current position and a shift in the read/write head to either the left or right $(L/R)$ by one position. The \textit{complete configuration} of the Turing machine can be seen as the combined state of the state registry and the symbols on the tape. And we can view the functioning of the Turing machine as set of transitions between these different \textit{complete configurations} $\{q,x\} \in Q \times \Sigma$. The machine halts with the state (registry) of machine in $\{q_F\} \in F$ and the output on the tape.

A simple example of a general substitution in Turing machines is through the use of a  universal Turing machine $\mathcal{U}$ which can simulate an arbitrary Turing machine $\mathcal{M}$ on arbitrary input by reading both the description of the machine to be simulated as well as the input to that machine from its own tape. Thus a universal Turing machine $\mathcal{U}$ can also be defined by a similar 7-tuple with an expanded set of input symbols $\Sigma^{\mathcal{U}}$ (to accept original inputs and description of $\mathcal{M}$) and states space of the state registry $Q^{\mathcal{U}}$. When the machine $\mathcal{U}$ halts, the symbols on the tape should correspond to the same output generated by $\mathcal{M}$ upon halting. The complete configuration of $\mathcal{U}$ is given by $\{q^{\prime},x^{\prime} \} \in Q^{\mathcal{U}} \times \Sigma^{\mathcal{U}}$. Since $\mathcal{U}$ will produce the same output on the tape as $\mathcal{M}$ upon halting for the same inputs, albeit through a different set of registry states from the space $Q^{\mathcal{U}}$ and transition function $\delta^{\mathcal{U}}$, we can construct a substitution using the concatenated state $\{q_0 q_1 q_2...q_F\}$ and $\{q_0^{\mathcal{U}}  q_1^{\mathcal{U}} q_2^{\mathcal{U}} ....q_F^{\mathcal{U}} \}$ as prediction data of $\mathcal{M}$ and $\mathcal{U}$ respectively, and the final output on the tape when the machine halts as the inference data.

Now imagine a Level-1 functionalist theory of consciousness $\mathcal{T}$. We can see that for any input-output pair $(x,o)$, we can write the corresponding Turing machine realization as $(q_0,x) \rightarrow (\bar{q}=\{ q_1 q_2....q_F \},o)$ and $(q_0^{\mathcal{U}},x \times \mathcal{X}_{\mathcal{M}}) \rightarrow (\bar{q}^{\mathcal{U}}=\{ q_1^{\mathcal{U}} q_2^{\mathcal{U}} ....q_F^{\mathcal{U}} \},o)$ in $\mathcal{M}$ and $\mathcal{U}$ respectively, where $\mathcal{X}_{\mathcal{M}}$ is the description of $\mathcal{M}$ as input to $\mathcal{U}$. We can rewrite the initial state of $\mathcal{U}$ by combining $\mathcal{X}_{\mathcal{M}}$ into the initial state $q_0^\mathcal{U}$ to produce $q_0^{\mathcal{U}*}=(q_0^{\mathcal{U}}, \mathcal{X}_{\mathcal{M}})$ and thus maintain the same `input' $x$. Plugging this back in for the machine $\mathcal{U}$, we have  $(q_0^{\mathcal{U}*},x) \rightarrow (\bar{q}^{\mathcal{U}}=\{ q_1^{\mathcal{U}} \times....q_F^{\mathcal{U}} \},o)$  Even though $\{q \}$ and $\{q^{\mathcal{U}} \}$ are elements in different spaces and $\bar{q}$ and $\bar{q}^{\mathcal{U}}$ are of different lengths ($\mathcal{M}$ and $\mathcal{U}$ need not produce the output in the same order of time-complexity), their topological functional relationship remains the same across configurations of the two Turing machines. Thus for every $\bar{q} \in \{ \bar{q} \}$ with $(q_0,x) \rightarrow (\bar{q},o)$, there is a $\bar{q}^{\mathcal{U}} \in \{ \bar{q}^{\mathcal{U}}\}$ with $(q_0^{\mathcal{U}*},x) \rightarrow (\bar{q}^{\mathcal{U}},o)$ across all input-output pairs $(x,o)$. This is isomorphic to the example of the hidden states in the RNN example used earlier in the section. The very same arguments can be extended here to show that though $\bar{q} \neq \bar{q}^{\mathcal{U}}$, we would have $pred_{\mathcal{T}}(\bar{q}) \cap pred_{\mathcal{T}}(\bar{q}^{\mathcal{U}}) \neq \emptyset$ while output remains the same. For such a theory of consciousness $\mathcal{T}$, the variation between Turing machine $\mathcal{M}$ and an universal Turing machine $\mathcal{U}$ is a Type-1 variation and does not imply pre-falsification. Both Turing machines here are universal models of computations capable of simulating any computable function and equivalent to other universal models like Lambda calculus, cellular automata, recursive functions, etc. It follows that \textbf{for a Level-1 functionalist theory $\mathcal{T}$ as defined above, a variation between any two universal models of computation will be a Type-1 variation with respect to $\mathcal{T}$ and will not imply it's pre-falsification.} Of course, as in the case of the RNN, we can construct a theory of consciousness $\mathcal{T}^{\prime}$ that is sensitive to Level-2 descriptions such that the variation from $\mathcal{M}$ to $\mathcal{U}$ is of Type-2 and $\mathcal{T}^{\prime}$ is pre-falsified by Theorem 1. The examples discussed in this section strongly hint at the non-existence of Type-2 variations for Level-1 functionalist theories of consciousness. We will prove that this is indeed the case in the next section.

\section{No Substitute for Level-1 Functionalism - A Formal Proof}
The authors of \cite{Walker} proposed the following idea - \textit{'For a theory to avoid the epistemic problems revealed by IIT under the isomorphic
transformation we introduce requires that no transformation or "substitution" exists that changes the prediction without affecting the inference. This, in turn, implies that beneath the specified level of inference, a mathematical theory of consciousness must be invariant with respect to any and all changes that leave the results from the inference procedure fixed. In other words, if you can make a change to the physical system that does not affect what will be used to infer the conscious state of the system, then such a change must not affect the prediction of the theory either.'} In this section, we will build on the examples from the previous section and provide a short formal proof (by contradiction) that no substitutions exist by showing no Type-2 variations exist for Level-1 functionalist theories (of consciousness). As a result, these cannot be pre-falsified by the substitution argument of \cite{Hoel}. We have briefly described Level-1 functional description in the previous section but will provide a more precise definition here. A Level-1 functionalist framework $\mathcal{T}_F$ is defined in terms of

\begin{itemize}
    \item \textbf{Functional states} $s^{\mathcal{T}_F} \in \mathcal{S}^{\mathcal{T}_F}$  which are defined mathematically in terms of their topological relationship to other functional states. They are thus independent of any particular encoding or representation.
    
    \item This topological relationship, which we call the \textbf{functional structure} $\mathcal{G}_F$ can be characterized as the map: $(s_{(t-1)}^{\mathcal{T}_F},x_t) \rightarrow (s_t^{\mathcal{T}_F},o_t) $ defined over all $(x_t,o_t)$. $x_t$ and $o_t$ are inputs and outputs at time $t$. Note that we can also combine $o_t$ into the functional state $s_t^{\mathcal{T}_F}$ and rewrite the map as $(s_{(t-1)}^{\mathcal{T}_F},x_t) \rightarrow s_t^{\mathcal{T}_F}$. If there are a collection of functional states $\{s \}$ and $\{ s^{\prime} \}$, and for every $s \in \{ s\}$, there is (atleast) a $s^{\prime} \in \{ s^{\prime} \}$ such that $(s_{(t-1)},x_t) \rightarrow (s_t,o_t)$ and $(s^{\prime}_{(t-1)},x_t) \rightarrow (s^{\prime}_t,o_t)$ over all $(x_t,o_t)$ pairs, then $\{s \}$ and $\{s^{\prime}\}$ share the same functional structure and the corresponding $s \in \{ s\}$ and $s^{\prime} \in \{ s^{\prime} \}$ are equivalent functional states.
    
    \item The result of the prediction correspondence $pred$ only depends upon the set of functional states $\{ s^{\mathcal{T}_F} \}$. If $s \in \{ s\}$ and $s^{\prime} \in \{ s^{\prime} \}$ are equivalent functional states, then $pred(s)=pred(s^{\prime})$ (or in a broader sense $pred(s) \cap pred(s^{\prime}) \neq \emptyset$). In general, the $pred$ function/correspondence from \cite{Hoel} defined for prediction data $o \in \mathcal{O}$ can be decomposed into a composition of two functions - an abstraction function $abs: \mathcal{O} \rightarrow \mathcal{S}^{\mathcal{T}_F}$ that maps prediction data in $\mathcal{O}$ to functional states in $\mathcal{S}^{\mathcal{T}_F}$. And a $pred^{\mathcal{T}_F}: \mathcal{S}^{\mathcal{T}_F} \rightarrow \mathcal{E} $ that maps functional states to prediction results in the experience space $\mathcal{E}$. Thus $pred(o)=pred^{\mathcal{T}_F}(abs(o))$.
\end{itemize}

This abstraction function $abs: \mathcal{O} \rightarrow \mathcal{S}^{\mathcal{T}_F}$ always exists since it can always be constructed for a Level-1 functionalist framework. In the prediction data $o_i$, let there be a particular physical representation of states $\{ h^{\mathcal{T}_F} \} \in \mathcal{H}^{\mathcal{T}_F} $. If $ \{ h^{\mathcal{T}_F} \}$ is such that for every input-output pair $(x_t,o_t)$, we have $(h_{(t-1)}^{\mathcal{T}_F},x_t) \rightarrow (h_t^{\mathcal{T}_F},o_t)$ and a $s_{(t-1)}^{\mathcal{T}_F} \in \{s^{\mathcal{T}_F} \}$ with $(s_{(t-1)}^{\mathcal{T}_F},x_t) \rightarrow (s_t^{\mathcal{T}_F},o_t)$, then we define $abs(h_{(t-1)}^{\mathcal{T}_F})=s_{(t-1)}^{\mathcal{T}_F}$, thereby constructing it into existence. By the very definition of a Level-1 functionalist theory of consciousness, $pred^{\mathcal{T}_F}: \mathcal{S}^{\mathcal{T}_F} \rightarrow \mathcal{E} $ also exists and would be built according to the particular framework. The output in general would be a non-invertible function of the functional state i.e. $o_t=g(s_t^{\mathcal{T}_F})$ is a coarse-graining over the functional states. In the case where $o_t$ has been concatenated to the functional state, the function $g$ simply traces out the non-output parts of the functional state. Note that since $g$ is assumed to be a non-invertible function, $g^{-1}$ need not exist and thus  $s_t^{\mathcal{T}_F}=g^{-1}(o_t)$ is not necessarily defined. If we view the output as corresponding to the inference data $o_r$, we have that $\{ s^{\mathcal{T}_F}\} \not\subseteq \{ o\}$ (i.e $o_i \not\subseteq o_r$) and thus we have not defined our Level-1 functionalist framework to be (pathologically) unfalsifiable as described by the conditions in \cite{Hoel}. 

We will briefly take a moment here to discuss the Reductio ad absurdum argument from \cite{Hoel}. The authors argue that for experiments in the natural sciences - \textit{``If there are two quantities of interest whose relation is to be modeled by a scientific theory, then in all reasonable cases there are two independent means of collecting information relevant to a test of the theory, one providing a dataset that is determined by the first quantity, and one providing a dataset that is determined by the second quantity.''} They explain this with an example of the relationship between temperature $T_0$ and it's relationship to the energy of microphysical states. They argue that in order to determine this for any particular model we would make two different measurements - one that would access the microphysical states and measure their kinetic energy (say) which would correspond to the prediction data ($o_m$) and the other would use a thermometer to obtain a dataset $o_{T_0}$ that replaces the inference dataset. They claim that \textit{`these independent means provide independent access to each of the two datasets in question'} and this \textit{'differs from the case of theories of consciousness considered here, wherein the physical system determines both datasets.'} We are not certain as to whether independent is used according to how it has been defined earlier in the paper. Nonetheless, we do not understand why in this example, the authors of \cite{Hoel} believe that the temperature measurement of the system $P$ using the thermometer does not depend upon the physical system $P$? The measurement obtained from the thermometer is in fact a calibrated value that provides a coarse-grained macroscopic description of the underlying microscopic kinetic energies (via a Maxwell-Boltzmann distribution of particle speeds). So if the authors identify no issues associated with the example they provided, then it should follow that for the sake of consistency, they will have no objections to the Level-1 functionalist framework defined above since the inference data/output is simply a coarse-grained result of the functional states achieved through an non-invertible function. We can now proceed further with our framework $\mathcal{T}_F$.

Since the functional states $ s^{\mathcal{T}_F} \in \mathcal{S}^{\mathcal{T}_F} $ are independent of any specific encoding, we will define the set $\{ \mathcal{H}_i^{\mathcal{T}_F} \}$ as the collection of different representations/encodings of Level-1 functional states that realize the same functional structure - $\{ \mathcal{H}_i^{\mathcal{T}_F} \} = \{ \mathcal{H}_1^{\mathcal{T}_F},\mathcal{H}_2^{\mathcal{T}_F},..., \mathcal{H}_N^{\mathcal{T}_F},...\}$. Each element $\mathcal{H}_j^{\mathcal{T}_F} \in \{ \mathcal{H}_i^{\mathcal{T}_F} \}$ is itself a collection of states $\{ h_i^{\mathcal{T}_F} \}$ that corresponds to a very particular physical representation of the individual functional states that maintain the same functional structure as any other $\mathcal{H}_k^{\mathcal{T}_F} \in \{ \mathcal{H}_i^{\mathcal{T}_F} \}$. Under this definition of $\{ \mathcal{H}_i^{\mathcal{T}_F} \}$, consider any pair of encodings $\mathcal{H}, \mathcal{H}^{\prime} \in \{ \mathcal{H}_i^{\mathcal{T}_F} \}$ (where we have dropped $\mathcal{T}_F$ to keep the equations cleaner). Since both $ \mathcal{H}$ and $\mathcal{H}^{\prime}$ have the same functional structure $\mathcal{G}_F$, by definition we have that for every $h^j, h^{j^{\prime}} \in \mathcal{H}$, there is a $h^k, h^{k^{\prime}} \in \mathcal{H}^{\prime}$, such that $(h_{(t-1)}^{j},x^t) \rightarrow (h_t^{j^{\prime}},o^t)$ and $(h_{(t-1)}^{k},x^t) \rightarrow (h_t^{k^{\prime}},o^t)$ for all input-output pairs $(x^t,o^t)$. This means that the different encodings $h^{j}$ and $h^{k}$ correspond to equivalent functional states. We thus have that for different encodings $\mathcal{H}$ and $\mathcal{H}^{\prime}$ that share the same functional structure, we can write using the $abs$ and $pred^{\mathcal{T}_F}$ function from before, $s=abs(h^{j}) \in \{s\}$, $s^{\prime}=abs(h^{k}) \in \{s^{\prime} \}$ where $pred^{\mathcal{T}_F}(s)=pred^{\mathcal{T}_F}(s^{\prime})$ (or $pred^{\mathcal{T}_F}(s) \cap pred^{\mathcal{T}_F}(s^{\prime}) \neq \emptyset$).

We will now prove by contradiction that Type-2 variations do not exist for Level-1 functionalist frameworks. Let us say that there is a pre-falsification achieved through the existence of Type-2 variation for the theory $\mathcal{T}_F$. This means that there is a $v: P \rightarrow P $ such that the inference data $o_r$ is kept constant while the prediction data changes from $o_i$ to $o_i^{\prime}=obs(v(p)) \neq o_i$, and the corresponding predictions do not overlap either i.e. we have $pred(o)\cap pred(o^{\prime})= \emptyset$ for $v$. From discussions in \cite{Hoel}, we take the prediction $o_i$ and inference $o_r$ data to correspond to particular encodings of the functional states and outputs respectively in order to be able to apply the functionalist model of interest. This means that $o_i \equiv \mathcal{H}$, $obs(v(p))=o_i^{\prime} \equiv \mathcal{H}^{\prime}$ and $o_r = \{ o\}$ (outputs). As $o_i \neq o_i^{\prime}$, we have that the corresponding encodings $\mathcal{H}$ and $\mathcal{H}^{\prime}$ are different in some measurable way. Since $v$ is a variation that maintains the inference data for the same experiments (i.e. maintains all input-output pairs across the variation), that would mean that for every $h^{j} \in \mathcal{H}$, there is a $h^{k} \in \mathcal{H}^{\prime}$ such that $(h_{(t-1)}^{j},x^t) \rightarrow (h_t^{j^{\prime}},o^t)$ and $(h_{(t-1)}^{k},x^t) \rightarrow (h_t^{k^{\prime}},o^t)$ for all input-output pairs $(x^t,o^t)$. This means that both $\mathcal{H}$ and  $\mathcal{H}^{\prime}$ have the same functional structure by definition (and are elements of $\{ \mathcal{H}_i^{\mathcal{T}_F} \})$. This would also mean that the $h^{j}$ and $h^{k}$ from above correspond to equivalent functional states (say $s$ and $s^{\prime}$) such that $s=abs(h^{j}) \in \{s\}$, $s^{\prime}=abs(h^{k}) \in \{s^{\prime} \}$ with $pred^{\mathcal{T}_F}(s)=pred^{\mathcal{T}_F}(s^{\prime})$ (or $pred^{\mathcal{T}_F}(s) \cap pred^{\mathcal{T}_F}(s^{\prime}) \neq \emptyset$).

The $pred$ function for $\mathcal{T}_F$ was defined as a composition of two functions $pred^{\mathcal{T}_F}$ and $abs$ functions. Thus for any (arbitrary) single prediction $pred(o)= pred( h^{j})= pred^{\mathcal{T}_F}(abs( h^{j}))$ and the corresponding prediction on the data obtained after the variation - $pred(o^{\prime})= pred( h^{k})= pred^{\mathcal{T}_F}(abs( h^t_{k}))$. Since we know that for any $h^{j} \in \mathcal{H}$ and $h^{k} \in \mathcal{H}^{\prime}$ that share the same functional relationships, we have $pred^{\mathcal{T}_F}(abs( h^{j})) = pred^{\mathcal{T}_F}(abs( h^{k})) $. This means that for any single prediction $pred(o)=pred(o^{\prime})$ (or rather $pred(o) \cap pred(o^{\prime}) \neq \emptyset$). Since the choice of $ h^{j}$ was arbitrary, we have $pred(o)\cap pred(o^{\prime}) \neq \emptyset$ over the prediction dataset. But we began with the assumption that $v$ is a Type-2 variation with $pred(o)\cap pred(o^{\prime})= \emptyset$. This is clearly a contradiction and thus no such Type-2 variation $v$ exists for a Level-1 functionalist theory $\mathcal{T}_F$. Thus prediction and inference data are independent via Type-1 variations and there no substitutions. Hence we have proved that such functionalist frameworks cannot be pre-falsified by the substitution argument presented in \cite{Hoel}. While the \textit{unfolding argument} can be seen as a special case of the \textit{substitution argument}, the above proof indicates that the claim about Level-1 functionalist theories \textit{like GWT, HOT and PP frameworks} being unaffected \cite{Doerig} by the unfolding (and now substitution) argument still hold. While we have shown that no substitutions exist for Level-1 functionalist theories, there is an underlying question about the status of Level-2 functionalist frameworks and whether some of the current leading frameworks fall under that category. This will be the focus of our next section.

\section{Level-1 vs Level-2 Functionalism}
One might be tempted to characterize frameworks that depend on the CSA descriptions as Level-2 functionalist theories, since the different states in each individual representation is defined in terms of it's relationship to other states under the same representation. (The author is not sure what the Level-1 structure corresponds to if \textit{functional structure} is defined at Level-2.) However with functional structure defined based on Level-2 state encoding, we can see how one can look at the different encodings of the same Level-1 \textit{functional} state and mistake them to be different Level-1 states - for eg: both $111^1$ and $001^2$ in Fig.(\ref{Walker_fig}) correspond to same Level-1 state $E$ in the context of the entire functional structure. However we see that $111^1 \rightarrow 011^1$ and $001^2 \rightarrow 101^2$ (and $111^2 \rightarrow 000^2$) which can lead to an erroneous conclusion that $111^1$ and $001^2$ correspond to different Level-1 functional states. The important thing to note is the superscript $1$ and $2$ over the states that correspond to different Level-2 descriptions and we must view these state encodings within the context of the \textbf{entire} CSA representation i.e. the overall functional structure to determine if they correspond to the same Level-1 functional state. If we fail to do that and distinguish between $111^1$ and $001^2$, then we are capturing something else about the states - like the difference in architecture - feedback vs no-feedback that is reflected in the $\phi$ calculations for IIT \cite{Walker}. We would then be more accurate in describing frameworks that differentiate based on these Level-2 descriptions by the specific factor/constraint that produced the difference in representations in the 1st place (for eg: it would be more accurate to call the framework that distinguishes between CSA representations 1 and 2 in Fig.(\ref{Walker_fig}) as a \textit{feedback theory of consciousness}). 

This is the case in \cite{Kleiner} where the author discusses a specific type of functionalism called \textit{machine state functionalism} where \textit{`any creature with a mind can be regarded as a Turing machine (an idealized finite state digital computer), whose operation can be fully specified by a set of instructions (a ``machine table'' or program) each having the form - If the machine is in state} $S_i$\textit{, and receives input} $I_j$\textit{, it will go into state} $S_k$ \textit{and produce output} $O_l$ \textit{(for a finite number of states, inputs and outputs)'} \cite{Levin}. The \textit{machine table} as defined above would be a collection of these instructions of the form $(S_i,I_j) \rightarrow (S_k,O_l)$ which is similar to how we characterized functional structure here. However this particular definition of machine state functionalism is adopted at a Level-2 description in \cite{Kleiner}, which leads to \textit{Level-2 functionalism} (as discussed above). As a result, changes to the automaton from feedback to feedforward architectures that is reflected in the Level-2 state encodings are viewed as changes to the functional structure. Since the Krohn-Rhodes decomposition \cite{KR} used to achieve the transformation between the two state representations assumes the preservation of the underlying (Level-1) functional structure, there is now confusion as to what actually constitutes functional structure. We could have different definitions for Level-1 and Level-2 functional structures, where the Level-1 structure is preserved across different Level-2 representations (arising from different physical realizations) and it is not as per the definition of the Level-2 structure. While this would be coherent in principle, results and claims now trivially become about which definitions have been adopted. In order to avoid this, it is important that we adopt and practice the use of terms in a consistent manner by determining \textit{which of the two definitions of functional structure is closer to what we originally intended to capture.} This difference between defining functional structure at Level-1 vs Level-2 descriptions also corresponds to the unresolved debate between \textit{role} and \textit{realizer} functionalism \cite{Levin}. Given how functionalism and functional states are conceptualized to be independent of the realization details plus given the arguments in favor of role functionalism \cite{Levin}, the author leans more towards a Level-1 definition as being closer to what functional structure intended to characterize. 

Consequently, claims made about Global Workspace theory (GWT), Global Neuronal Workspace theory (GNWT) along with other functionalist frameworks in \cite{Kleiner} based on \textbf{Lemma 2.9} do not follow unless one adopts a Level-2 definition of functional structure (According to \textbf{Definition 2.8} in \cite{Kleiner}, a Level-1 functionalist picture as defined in the previous section will not satisfy Lemma 2.9 for the same reasons it is immune to the substitution argument). While the description of GNWT from \cite{Stanislas} that is used in \cite{Kleiner} highlight a very architecture specific Level-2 type description (which could motivate one to view GNWT as a Level-2 functionalist framework), the authors of \cite{Stanislas} clearly state in their paper that this particular description is \textit{``from a neuronal architecture standpoint.''} Furthermore they start this description by pointing out that the original GNWT relies on the following main assumptions - \textit{``that conscious access is global information availability: what we subjectively experience as conscious access is the selection, amplification and global broadcasting, to many distant areas, of a single piece of information selected for its salience or relevance to current goals''} - which is more of a Level-1 description that is independent of how selection, amplification and global broadcasting is realized. The authors in \cite{Doerig} make a similar point with respect to Global Workspace, Higher-Order Thought (HOT) and Predictive Processing (PP) theories which they characterize as Level-1 functionalist frameworks - \textit{``The unfolding argument does not apply to these theories because they propose that systems are conscious in so far as they implement the right kind of function–independently of the causal structure. Of course, these theories are usually couched in terms of recurrent or top-down processing, or other seemingly causal-structure terminology, but they can be formulated in other kinds of networks too''} and provide a feedforward toy model of GWT to further strengthen their case. We can leave the status of whether these leading theories of consciousness are Level-1 or Level-2 functionalist frameworks as open, but reiterate that claims of pre-falsification only apply if their functional structure is couched in Level-2 descriptions. Since we can always construct an abstraction function $abs$ to map from the Level-2 to Level-1 description, it is possible to construct a Level-1 version of a Level-2 framework and avoid pre-falsification via substitution. On the other hand, one could argue that we can always construct a specific Level-2 realization of a Level-1 theory (by constraining aspects of the architecture, use of neuronal units, etc), which would then be fallible to the substitution argument. This would at best only pre-falsify by substitution that specific Level-2 version and not the underlying Level-1 framework, and at worst bring us back to square one on whether or not the Level-2 description is a functionalist framework to begin with. Finally, if we used the definition of functional structure from \cite{Kleiner} and followed the results of \cite{Walker}, it seems like we would have both functionalist and causal structure being defined on the same Level-2 (CSA) descriptions which appears to be contradictory to how many view both and would only lead to further confusion. It would be important to pin this down in a coherent and consistent manner before they lead to further debates over pre-falsification. 


\section{Summary \& Conclusion}
The work presented in \cite{Hoel} represents a good first step in formalizing the techniques used in the science of consciousness. The corresponding substitution arguments would have important implications in the field of consciousness by pointing at fundamental problems in the manner in which experiments are conducted that would pre-falsify many of the current major frameworks of consciousness. However we find that claims like - \textit{``We come to a surprising conclusion: a widespread experimental assumption implies that most contemporary theories of consciousness are already falsified''} by the authors are currently unjustified once you take into account that \textit{independence} was defined to trivially imply pre-falsification. In this reply, we introduced a more complete definition of independence that allowed us to expand variations from the original paper into Type-1 and Type-2. We then redefined Theorem 3.10 from \cite{Hoel} using Type-2 variations and then explored substitutions in the case of neural networks, state machines with and without feedback and Turing machines. We showed that for these particular example cases, the substitutions of interest were not of Type-2 with respect to a Level-1 functionalist framework and hence does not entail pre-falsification. We then presented a formal proof of the non-existence of Type-2 variations for Level-1 functionalist frameworks of consciousness and completed the reply with a discussion of Level 1 and 2 functionalist theories and where contemporary theories of consciousness fall into. Currently, we are uncertain as to whether or not most leading functionalist pictures of consciousness are of Level-1 or 2. What is more certain however is that they have not already been pre-falsified independent of that determination.

\bibliographystyle{unsrt}  


\end{document}